\documentclass[11pt,preprintnumbers,a4paper]{article}
\usepackage{jheppub}
\usepackage{bbm}
\usepackage{epsfig,bm}
\usepackage{graphics,graphicx,comment,subfigure}

\pdfoutput=1

\newcommand{\A}{\mathcal{A}}
\newcommand{\F}{\mathcal{F}}
\newcommand{\hA}{\hat{A}}

\newcommand{\zuv}{z_{\text{uv}}}
\newcommand{\UKK}{U_{\text{KK}}}
\newcommand{\Uuv}{U_{\text{uv}}}
\newcommand{\MKK}{M_{\text{KK}}}

\newcommand{\be}{\begin{equation}}
\newcommand{\ee}{\end{equation}}

\def\bea{\begin{align}}
\def\ena{\end{align}}

\def\nnb{\nonumber}
\def\Tr{\mbox{ Tr }}
\def\beqa{\begin{eqnarray}}
\def\enqa{\end{eqnarray}}
\def\nnb{\nonumber}
\def \Nc{N_c}

\def\XPT{$\chi$-PT}

\title{Long-distance properties of baryons in the Sakai-Sugimoto model}
\author{Aleksey Cherman and}
\emailAdd{a.cherman@damtp.cam.ac.uk}
\author{Takaaki Ishii}
\emailAdd{T.Ishii@damtp.cam.ac.uk}
\affiliation{%
Department of Applied Mathematics and Theoretical Physics,\\ 
University  of Cambridge, \\
Cambridge CB3 0WB,
UK
}

\abstract{%
We reconsider the realization of baryons in the Sakai-Sugimoto model.  In this theory, which is the gravity dual of a QCD-like theory, baryons appear as soliton solutions.  These solitons were approximated as flat-space instantons in previous studies.   However, with this approximation, it has been shown that one does not reproduce some model-independent predictions for the behavior of baryon electromagnetic form factors which are connected with long-range pion physics.  This made it appear that the long-range pion physics of baryons may be hidden in (intractable) $\alpha'$ corrections in the gravity dual.  In this paper, we study the long-range properties of baryons in the Sakai-Sugimoto model without relying on the flat-space instanton approximation.  The solution we obtain gives the correct result for the model-independent ratio of form factors, implying that the model captures the expected infrared properties of baryons without the need to go beyond the leading order in the $\alpha'$ expansion.
}

\begin{document}

\maketitle
%%%%%%%%%%%
\section{Introduction}
\label{sec:introduction}
%%%%%%%%%%%

Thanks to the development of gauge-gravity duality\cite{Maldacena:1997re,Witten:1998xy,Gubser:1998bc}, we can use gravity duals as powerful computational tools for analyzing strongly-coupled  QCD-like theories. Perhaps the most prominent QCD-like theory with a gravity dual is the Sakai-Sugimoto model\cite{Sakai:2004cn,Sakai:2005yt}. The model has a spectrum of weakly-interacting mesons with the same quantum numbers as in QCD, and baryons can be thought of as solitons of the meson fields, just as one would expect in large $N_{c}$ QCD\cite{tHooft:1973jz,Witten:1979kh}.  The price paid for the tractability of the model is that in addition to the fields and particles that are contained in large $N_{c}$ QCD, there are also many other particles at the same mass scale as the particles with QCD quantum numbers. These extra fields conspire to make the 't Hooft coupling $\lambda = g_{YM}^{2} N_{c}$ a tunable control parameter of the model, which is necessary for the existence of a classical gravity dual.  Since the Sakai-Sugimoto model contains a great many of the ingredients of QCD, it is hoped that for many observables the model will give at least qualitatively accurate predictions.  Of course, this begs the questions of which observables in QCD have a good match in the Sakai-Sugimoto model, and how to define what counts as a good match.

At least one setting where such questions should have sharp answers is for low-energy QCD observables, where there are predictions from chiral perturbation theory (\XPT).   For such observables it is known precisely what one should expect from QCD, and sharp comparisons with the Sakai-Sugimoto model are possible. The question we focus on in this paper is whether the predictions of \XPT\ for baryons are satisfied by the Sakai-Sugimoto model.   Since the Sakai-Sugimoto model has spontaneous chiral symmetry breaking, and \XPT\ is simply a systematic way to work out the  the consequences of symmetry breaking for low-energy observables, this may seem like a foregone conclusion.  Things are not so simple, however.  As an effective field theory (EFT), \XPT\ assumes that the low-energy constants appearing in the derivative expansion are of natural sizes.  The Sakai-Sugimoto contains the extra parameter $\lambda$ compared to QCD, and it could be that some low-energy constants are suppressed by powers of $\lambda$.  If this happens then there will be low-energy observables that behave qualitatively differently between QCD and the Sakai-Sugimoto model.  Fortunately, no observables that disagree with the predictions of \XPT\ have been found in the meson sector. But in the baryon sector\footnote{For studies of baryons in the Sakai-Sugimoto model, see e.~g.~Refs.~\cite{Hong:2007kx,Hata:2007mb,Hong:2007ay,Hong:2007dq,Park:2008sp,Hashimoto:2008zw,Kim:2008pw}. }, things are less encouraging, and observables which appear to be afflicted by this issue have been previously identified in Ref.~\cite{Cherman:2009gb}.  The observables discussed in Ref.~\cite{Cherman:2009gb} are the long-distance limits of the Fourier transforms of the electric and magnetic form factors of the proton, which encode the response of the proton to soft photon probes\footnote{Unlike other probes of the long-distance behavior of baryons such as charge radii, which are related to the moments of form factors, the Fourier transforms of the form factors remain finite in the chiral limit.  }.  The leading long-distance behavior of these observables is tightly constrained by $\chi$-PT, and so must satisfy some model-independent relations in QCD-like theories.

\begin{figure}
        \label{fig:FormFactorDiagrams}
        \centering
        \subfigure[Isoscalar]{
            \label{ChiPT:scalar}
            \includegraphics[width=1.5in]{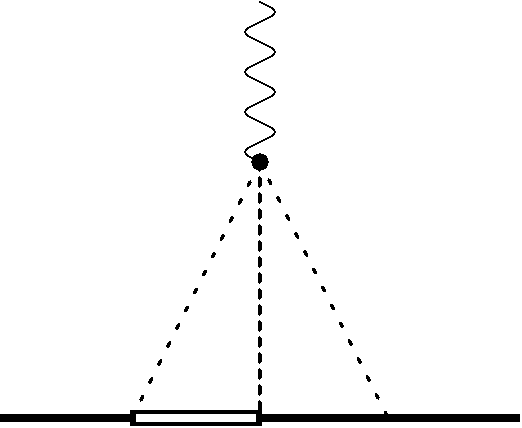}
        }
        \subfigure[Isovector]{
            \label{ChiPT:vector}
            \includegraphics[width=1.5in]{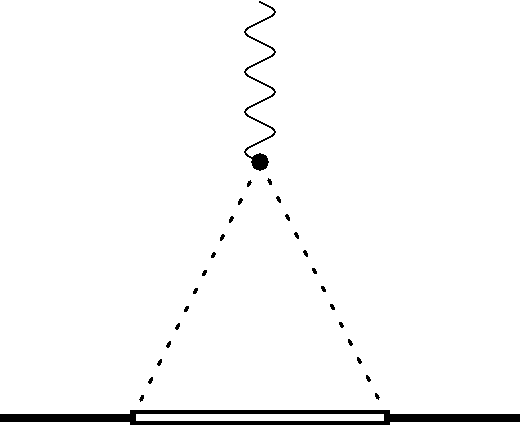}
        }
    \caption{Some representative diagrams that contribute to the long-distance behavior of form factors in large $\Nc$ $\chi$-PT.  Fig.~\ref{ChiPT:scalar} depicts a leading contribution to the nucleon isoscalar form factor, which is dominated by a three-pion interaction from the anomaly.  Fig.~\ref{ChiPT:vector} depicts a leading contribution to the isovector nucleon form factor, which is dominated by a two-pion interaction.  Since the $\Delta$ baryon and the nucleons are degenerate at large $N_{c}$, the intermediate states can be either nucleons or $\Delta$s.  The $\Delta$s are shown as double lines above. }
\end{figure}

Specifically, Ref.~\cite{Cherman:2009gb} argued that in the chiral limit $m_{\pi}=0$, these position-space form factors must obey the relation\footnote{The chiral limit and the large $N_{c}$ limit do not commute.  In this paper we assume that the large $N_{c}$ limit is taken first.   While we work at $m_{\pi}=0$ throughout, it would also be interesting to explore what happens at finite $m_{\pi}$, as was done in the Pomarol-Wulzer model in Ref.~\cite{Domenech:2010aq}.}
\begin{align}
\label{eq:MagicRatio}
\lim_{r\to\infty} r^{2} \frac{\tilde{G}_{E}^{I=0}\tilde{G}_{E}^{I=1}}{\tilde{G}_{M}^{I=0}\tilde{G}_{M}^{I=1}} = 18,
\end{align}
where the $\tilde{G}$'s are isoscalar/isovector electric/magnetic form factors of the proton in position space.  These observables are just the Fourier transforms of the usual momentum-space form factors $G^{I=0,1}_{E,M}$, and are defined as
\begin{align}
\label{eq:FormFactorDefinition}
\tilde{G}_E^{I=0}(r) &= \frac{1}{4\pi} \int~d\Omega\langle p\uparrow|J_{I=0}^{0}|p\uparrow\rangle   \\
\tilde{G}_M^{I=0}(r) &= \frac{1}{4\pi} \int~d\Omega{1\over2}\varepsilon_{i j 3}\langle p \uparrow|x_{i} J_{I=0}^{j}|p\uparrow\rangle \\
\tilde{G}_E^{I=1}(r) &= \frac{1}{4\pi} \int~d\Omega\langle p\uparrow|J_{I=1}^{\mu = 0, a = 3} | p \uparrow\rangle  \\
\tilde{G}_M^{I=1}(r) &= \frac{1}{4\pi} \int~d\Omega{1\over2} \varepsilon_{i j 3}\langle p\uparrow| x_{i} J_{I=1}^{\mu=j, a=3} | p \uparrow\rangle 
\end{align}
where $J_{I=0}^{\mu}, J^{\mu, a}_{I=1}$ are the isoscalar and isovector currents, and the matrix elements are with respect to a proton in a spin `up' state to be specific.  In the chiral limit the pions are massless, and so give the leading contribution to the long-distance properties of the proton. The long-range behavior of these form factors is determined by the hadronic diagrams in Fig.~\ref{fig:FormFactorDiagrams}. Note that these diagrams are not suppressed in the large-$N_c$ limit despite seeming to contain a pion loop, because meson-baryon coupling constants scale with $N_{c}$\cite{Cohen:1989qk,Kiritsis:1988qz,Arnold:1990fp}.

The relation in Eq.~\eqref{eq:MagicRatio} is known to be satisfied by any chiral soliton model of baryons, such as the Skyrme model and its many generalizations. It is also satisfied by the Pomarol-Wulzer bottom-up holographic model of QCD\cite{Pomarol:2007kr,Pomarol:2008aa,Panico:2008it}.

%%%%%%%%%%%%%%%%%%%%%
%%%%%%%%%%%%%%%%%%%%%

The surprising conclusion of Ref.~\cite{Cherman:2009gb}, however, was that baryons in the Sakai-Sugimoto model did not appear to obey the model-independent relation. This conclusion was based on the standard treatment of baryons as holographic flat-space instantons in the Sakai-Sugimoto model \cite{Hata:2007mb,Hashimoto:2008zw}.   The flat-space instanton approximation is motivated by the large $\lambda$ limit, and appears to imply that the pion cloud around the nucleon is absent at large $\lambda$.   Instead, the flat-space instanton approximation suggests that the leading long-distance contribution to the form factors of the proton in the Sakai-Sugimoto model is mediated by $\rho$ mesons.  As a result, the long-distance behavior of the form factors is very different compared to what would expect from \XPT.    The large $\lambda$ limit is mapped to the $\alpha'$ expansions in the dual string theory, and as was discussed in Ref.~\cite{Cherman:2009gb}, the physics of the pion-nucleon interactions is then apparently hidden in the $\alpha'$ corrections in the string theory, which are not a calculable part of the model.   These considerations suggest that the large distance and the large $\lambda$ limits do not commute, so that if the large $\lambda$ limit is taken first the long-distance physics is completely different than in QCD.   So if the standard treatment in the literature is correct, then for this class of observables there are quite sharp qualitative differences between the Sakai-Sugimoto model and QCD.  Our aim is to  investigate whether this is indeed the case.

In this paper, we reexamine the calculations of the long-distance properties of baryons in the Sakai-Sugimoto model.  In contrast to previous treatments of the model, we do not start by assuming the flat-space instanton approximation, and directly solve for the long-distance behavior of the holographic soliton associated with the baryon in the dual field theory.  We then show that the structure of this long-distance solution is such that Eq.~\eqref{eq:MagicRatio} is in fact \emph{satisfied} by the baryons in the Sakai-Sugimoto model.  Our results are obtained directly from the leading terms in the action in the $\alpha'$ expansion.  So contrary to the pessimistic story above, the long-distance pion-nucleon physics is not hidden in the $\alpha'$ corrections to the Sakai-Sugimoto model.  It is present in the Sakai-Sugimoto model at leading order in the $\alpha'$ expansion.

This paper is organized as follows.  In Sec.~\ref{sec:SSmodel} we give an overview of the Sakai-Sugimoto model, describing the dual gravitational theory.  In this overview we emphasize the importance of introducing a UV cutoff in the gravity dual, which is not always appreciated.  We also describe the self-consistent ansatz for the soliton fields which is appropriate for investigating nucleons. We solve the equations of motion of the soliton in a $1/r$ expansion in Sec.~\ref{sec:LargeR}.  These solutions are turned into predictions for the long-distance behavior of the form factors using the standard machinery of collective coordinate quantization. The form factors are seen to obey Eq.~\eqref{eq:MagicRatio}. In Sec.~\ref{sec:discussion}, we make a few brief remarks about some possible issues in the previous treatments of baryons in the literature, which may help explain why the results we obtain in this paper were not obtained in previous analyses. Finally, we outline some possible directions for future work.

%%%%%%%%%%%
\section{The Sakai-Sugimoto model}
\label{sec:SSmodel}
%%%%%%%%%%%
\subsection{Brane construction}
The Sakai-Sugimoto model describes the strong-coupling physics associated with a system of intersecting $N_{c}$ $D4$ branes and $N_{f}$ $D8,\overline{D8}$ branes in type IIA string theory.  It is assumed that $N_{f}\ll N_{c}$.  At weak coupling, an open string picture is appropriate, and gluons can be seen to arise as open strings with both ends on the $N_{c}$ $D4$ branes, while quarks arise as open strings with one end on the $N_{c}$ color branes and another end on the $N_{f}$ flavor branes.  The $x_{4}$ direction is chosen to be a circle, with anti-periodic boundary conditions for fermions, which breaks supersymmetry.   The 4D field theory lives on the intersection of the world volumes of the color D4 branes and the flavor D8 branes.  Finally, baryon operators appear as  $D4$ branes wrapping the $x_{6}, \ldots, x_{9}$ directions.  The brane configuration with $N_{B}$ baryons is summarized in Table~\ref{table:BraneConfiguration}.
\begin{align}
\label{table:BraneConfiguration}
\begin{array}{c|cccccccccc}
   & x_0 & x_1 & x_2 & x_3 & (x_4) & x_5 & x_6 & x_7 & x_8 & x_9\\ \hline
N_{c}\, \mbox{D4} & \times & \times & \times & \times & \times & &  &  & & \\
N_{f}\, \mbox{D8},\overline{\mbox{D8}} & \times & \times & \times & \times &   & \times
& \times & \times &  \times &\times   \\
\hline
N_{B}\, \mbox{D4}& \times &  &  &  &   & 
& \times & \times &  \times &\times   \\
\end{array}
\end{align}
We will be interested only in configurations with $N_{B}=1$ in this paper.  Ref.~\cite{Sakai:2004cn,Sakai:2005yt} showed that despite the breaking of supersymmetry by the boundary conditions on the $S^{1}$, this brane configuration is stable, at least to leading order in a  $N_{f}/N_{c}$ expansion.

Let us briefly review the well-known connection of this brane construction to QCD\cite{Witten:1998zw,Sakai:2004cn,Sakai:2005yt}.  Because of the supersymmetry-breaking  boundary conditions on $S^{1}$, all fields except the gluons and quarks pick up masses of order $\MKK \sim 1/R_{S^{1}}$.   The 5D $SU(N_{c})$ gauge theory living on the $D4$ branes has a dimensionful 't Hooft coupling $\lambda_{5D}$.  The dimensionless parameter $\lambda = \lambda_{5D} \MKK$ has an interpretation as the 't Hooft coupling of a 4D theory evaluated at the scale $\MKK$.   In addition to massless gluons, the 4D theory has an infinite number of KK modes.  Then one can consider the two limits for the coupling at the scale $\MKK$:  $\lambda \to 0$ and $\lambda \to \infty$. If $\lambda$ is small at the KK scale, the KK modes become heavy and decouple from the dynamics at low energies (compared to $\MKK$).  The only massless fields are gluons and fundamental quarks, and the low-energy theory is just QCD, which confines at some dynamically-generated scale $\Lambda_{QCD} \ll \MKK$.  The brane system can thus be viewed as a particular UV completion of QCD at the scale $\MKK$.

On the other hand, one can also consider the opposite limit, with $\lambda$ large at the KK scale.  Now the Kaluza-Klein tower associated with the $x_{4}$ direction does not decouple, and the theory does not flow to pure QCD in the infrared.  However, if both $N_{c}$ and $\lambda$ are large, there is a dual description of the gauge theory in terms of gravity.  So at the cost of losing a sharp connection to the physics of pure QCD, one can profitably use the holographic description to study the strongly-coupled theory, which still turns out to share some of the most important features of QCD.   This is morally similar to working with the strong-coupling limit of a lattice gauge theory, which also allows many simplifications, at the expense of losing a sharp connection to the continuum field theory of interest.  The large $N_{c}$ and large $\lambda$ limits in the gauge theory are necessary to suppress $g_{s}$ and $\alpha'$ corrections in the dual string theory, which then reduces to classical gravity on a weakly-curved background.  With the optimistic assumption that the $\lambda \ll 1$ and $\lambda \gg 1$ limits of the theory are smoothly connected, without any phase transitions at some intermediate 't Hooft coupling, it may be hoped that calculations using the holographic theory will give results that are qualitatively similar to calculations in  QCD\footnote{For some interesting recent developments in this direction at finite temperature, see Ref.~\cite{Mandal:2011ws}.}.  With this hope in mind, the strong-coupling theory is often referred to as `holographic QCD' in the literature.

In the large $N_{c}$ and $\lambda$ limits, the D4 branes are replaced by their near-horizon geometry, with appropriate boundary conditions along for the $x_{4}$ circle\cite{Witten:1998zw}.  So long as $N_{f} \ll N_{c}$, the $D8$ branes have no effect on the geometry, and can be treated in the probe approximation.  In this limit the Type IIA supergravity equations of motion are solved by the metric $G_{MN}$, dilaton $\Phi$, and RR 3-form $C_{3}$ fields given in the string frame by\cite{Witten:1998zw}
\begin{align}
ds^{2}_{9+1} &=\left(\frac{U}{R}\right)^{3/2}  (\eta_{\mu \nu} dx^{\mu}dx^{\nu}+ f(U) d\tau^{2}) +\left(\frac{R}{U} \right)^{3/2}\left[ \frac{dU^{2}}{f(U)} + U^{2} d\Omega_{4}^{2} \right] \nonumber \\
e^{\Phi} &= g_{s} \left(\frac{U}{R}\right)^{3/4} ,\;\; f(U) = 1-\frac{\UKK^{3}}{U^{3}}, \; \; F_{4} = d C_{3} = \frac{2\pi N_{c}}{V_{4}} \epsilon_{4} 
\end{align}
where $R^{3} = \pi g_{s} N_{c} l_{s}^{3}$, $\eta_{\mu \nu}$ is the flat Minkowski metric, and $\tau =x_{4}$ and has circumference $2\pi/M_{\text{KK}}$.  Finally, $M_{\text{KK}} = 3 U_{\text{KK}}^{1/2}/2R^{3/2}$, $d\Omega_{4}$ is the line element on $S^{4}$ which has a volume $V_{4}$, and $\epsilon_{4}$ is the volume form on $S^{4}$.  The energy scale of the dual field theory is  related to  $U \MKK^{2}$.  Modulo some caveats that will be discussed below, $U$ takes values in the range $(U_{\text{KK}}, \infty)$. The background is topologically $R^{1,3} \times I \times S^{1}\times S^{4}$~\cite{Hong:2007ay}, with  $R^{1,3} \times S^{1} \times S^{4}$ fibered over the interval $I$, which is parametrized by $U$ above.  

Flavor fields arise from the D8 and $\overline{\textrm{D8}}$ branes\cite{Karch:2002sh}, which are immersed into the above background at the positions $\tau(U\to\infty) = \pm \delta\tau/2$ respectively.  There are $U(2)$ flavor gauge fields living on the D8 and  $\overline{\textrm{D8}}$  branes, which using the AdS/CFT dictionary are associated to sources for the currents of the $U(2)_{L} \times U(2)_{R}$ chiral symmetry of the dual field theory.  The embedding function $\tau(U)$ is determined from the equations of motion of the D8 branes, and the D8 branes turn out to connect to each other at $U=\UKK$.  This gives a beautiful geometric realization of spontaneous chiral symmetry breaking, as discussed in detail in Refs.~\cite{Sakai:2004cn,Sakai:2005yt}.

%%%%%%%%%%%%
\subsection{Region of validity of the supergravity approximation}
%%%%%%%%%%%%
To understand the region of validity of the supergravity approximation, some useful relations between the stringy parameters $g_{s},\MKK,\UKK,R, l_{s}$ and the field theory parameters $\lambda, N_{c}$ are 
\begin{align}
g_{s} = \frac{\lambda}{2\pi N_{c} \MKK l_{s}}, \;\; R^{3} = \frac{\lambda l_{s}^{2}}{2M_{\text{KK}}}, \;\; \UKK = \frac{2}{9} \lambda \MKK l_{s}^{2}.
\end{align} 
It turns out that the action depends on $l_{s}$ only through an overall normalization, and this can be used to set $\UKK = 1/\MKK$ without loss of generality.  Then one obtains the very useful relation $\frac{2}{9} \MKK^{2} l_{s}^{2} = \lambda^{-1}$.  The Type IIA supergravity description is valid provided that 1) all curvature invariants are small compared to the string scale, so that $\alpha'$ corrections can be neglected, and 2) the effective string coupling, which is controlled by the dilaton, remains small, so that string loop corrections can neglected\footnote{Once the dilaton becomes large, one can regain control of the theory by doing a lift to M-theory, so that the physics becomes describable in terms of 11-dimensional supergravity}.  In terms of the parameters of the dual field theory this translates into the constraint $1/\lambda \ll U \MKK  \ll N_{c}^{4/3}/\lambda$~\cite{Itzhaki:1998dd}\footnote{Fits to meson-sector data imply that for $N_{c} =3$, $\lambda \sim 17$, so that $\alpha'$ and $g_{s}$ corrections might be important.  From a formal perspective it is not obvious why the model works as well as it does in fitting the data.}.  In this case since the range of $U$ is bounded from below by $\UKK$, the lower limit above is not dangerous so long as $\lambda$ is large, because then $\alpha'$ corrections to the background will be negligible.  For large enough $U$, however, type IIA supergravity will not be reliable because the effective string coupling becomes large.  The fact that the region of validity of Type IIA supergravity is energy-dependent can be traced to the field-theory fact that 5D YM theory is non-renormalizable.  New degrees of freedom, which are not included in Type IIA supergravity, become important in the ultraviolet limit.   

In addition to the above issues with the region of validity of type IIA supergravity, in order for the holographic description of the theory to be reliable, one must make sure that the effective action used to describe the physics of bulk fields living on the flavor branes remains reliable.   This latter matter turns out to be a subtle business for baryons, because the effective action describing the flavor fields breaks down if the fields vary rapidly on distance scales comparable to $1/\alpha'^{1/2}$.  Studies using the flat-space instanton approximation suggest that the size  of the cores of baryons in the Sakai-Sugimoto model shrink to the string scale, and hence $\alpha'$ corrections to the effective action of the flavor gauge fields may not be under firm theoretical control in the small-$U \MKK$ region; see the discussions in Refs.~\cite{Hata:2007mb,Hashimoto:2008zw,Hong:2007ay,Hong:2007kx,Hong:2007dq,Hashimoto:2010je}.  Here we will be concerned with the low-energy properties of baryons, which the usual effective field theory ideology implies will only depend on such small-$U$ issues through a few `low-energy constants' such as {\it e.g.} the mass and moment of inertia of the holographic soliton configuration.  For the observables we will focus on here, it is expected that all such parameters would cancel out, and we will see that this is indeed the case with our approach.   

The treatment of the large $U$ region is much trickier. From the discussion above, it is clear that if the holographic description of the theory is to be reliable, one must make sure that the observables one is interested in receive no contributions from the region $U \MKK \gtrsim N_{c}^{4/3}/ \lambda$, where the supergravity theory is not under control.  To capture the long-range physics of baryons correctly we will see that it turns out to be critical to systematically sequester the large $U$ region by introducing a cutoff on $U$ at the intermediate stages of our calculations.  This brings us to the issue of holographic renormalization.

In holography, the UV divergences of field theory are traded for IR divergences of the on-shell action of the gravity dual due to the infinite extent of the extra holographic direction.  At the risk of some confusion we will stick with the conventions in the literature and refer to such IR holographic divergences as UV divergences.  The general setup of holography relates the large-$U$ behavior of bulk fields to the behavior of sources for various operators in the dual field theory.  In the well-understood examples of the AdS/CFT correspondence, it has been shown that one must impose a cutoff on the holographic coordinates at the intermediate stages of calculations in order to systematically remove volume divergences in the on-shell action of the holographic theory.  Once the theory has been regularized and any divergences subtracted by the addition of appropriate boundary counter terms, the cutoff can be - and should be - removed.  This procedure is known as holographic renormalization\cite{Skenderis:2002wp}.   

On general grounds, one expects that the same procedure is necessary for the Witten background\footnote{A systematic approach to holographic renormalization for non-conformal brane systems (such as the Witten background) has been developed only somewhat recently\cite{Kanitscheider:2008kd,Benincasa:2009ze,Papadimitriou:2010as,Papadimitriou:2011qb}.  To our knowledge these techniques have not yet been applied directly to the Sakai-Sugimoto model.}.  With this motivation, in our calculations we introduce a cutoff on $U$, so that $U \in (\UKK , U_{\mathrm{uv}}]$.     However, there is an important subtlety here which is not present in the tamer cases to which holographic renormalization is normally applied.  In view of the above observations on the domain of validity of the supergravity description of D4 branes, $U_{\mathrm{uv}}$ must be chosen such that $U_{\mathrm{uv}} \MKK \ll N_{c}^{4/3}/\lambda$ so that the behavior of the bulk theory near $U_{\max}$ can be trusted.  That is, even at the end of a calculation, in principle one should not strictly send $\Uuv \to \infty$. Instead $\Uuv$ must be taken larger than the other physical scales in the problem, but still remain small compared to $N_{c}^{4/3}/\lambda$, which is in principle a large but finite number in any given application.  Correlation functions in the field theory should then be calculated from the on-shell action of the regulated gravity theory, with the boundary values of bulk fields at $U=U_{\mathrm{uv}}$ acting as sources for field theory operators.  Well-behaved observables should have vanishing $U_{\mathrm{uv}}$ dependence as $U_{\mathrm{uv}}\MKK$ is taken to be large.   Indeed, it will turn out that working with a fixed cutoff $U_{\mathrm{uv}}$ and only removing the cutoff at the last step of the calculations plays a crucial role in the success of the Sakai-Sugimoto model in capturing the correct large distance physics of baryons. 

%%%%%%%%%%%%%%%
\subsection{Five-dimensional action}
\label{sec:FiveDim}
%%%%%%%%%%%%%%%
As long as the $U(N_{f})$ world-volume gauge fields $\A_{M}$, $M=0,\ldots,8$, on the D8 branes vary slowly compared to the string scale, their dynamics should be reliably described by a Yang-Mills-Chern-Simons action.  In the Sakai-Sugimoto model the field theory has an $SO(5)$ symmetry, under which the massless fields, which are the same ones that occur in QCD, are singlets.  We are interested in baryons with the quantum numbers of nucleons, which are $SO(5)$ singlets, and to this end we follow the approach in the literature of setting $\A_{5\cdots,8} =0$ and $\partial_{5,\ldots, 8} \A_{M}=0$.  It is also convenient to switch to a new holographic coordinate $z$ defined by
\begin{align}
U^{3} = \UKK^{3}+\UKK z^{2},
\end{align}
which takes values in $[-\zuv,+\zuv]$, with $\zuv \approx \Uuv^{3/2}/\UKK^{1/2}$.  With these choices, the Sakai-Sugimoto model reduces to a five-dimensional $U(N_f)$ Yang-Mills-Chern-Simons theory in a curved background.  The only dimensionful scale is $\MKK$, and we choose units such that $\MKK=1$.  The $\MKK$ dependence can always be restored by dimensional analysis.  The action is then\cite{Sakai:2004cn,Sakai:2005yt}
\begin{align}
\label{SS_action}
S_\mathrm{SS} =
-\kappa \int_{-z_\mathrm{uv}}^{z_\mathrm{uv}} d^4 x dz \,\mathrm{Tr}
\left[
\frac{1}{2} h(z) \mathcal{F}_{\mu\nu}^2 +  k(z) \mathcal{F}_{\mu z}^2
\right]
+ \frac{N_c}{24\pi^2} \int_\mathrm{M_5} \omega_5 + S_b.
\end{align}
Here  $\kappa = \lambda N_c/(216\pi^3)$,  and the functions $h(z) = (1+z^2)^{-1/3}$ and $k(z) = 1+z^2$ encode the metric of the curved background. The field strength is given by $\mathcal{F} = d \mathcal{A} - i \mathcal{A} \wedge \mathcal{A}$ and $\omega_5 = \mathrm{Tr} \left( \mathcal{A} \, \mathcal{F}^2 - i \mathcal{A}^3 \mathcal{F} /2 - \mathcal{A}^5 /10 \right)$ is the Chern-Simons five-form. In particular, we consider the case that $N_f=2$. 

Finally, $S_b$ stands for a set of boundary terms at $z=z_\mathrm{uv}$, which must be constructed from the boundary data in such a way that the total action is gauge and diffeomorphism invariant\cite{Skenderis:2002wp}.  The coefficients of some of these terms may be fixed by the requirement that the on-shell action evaluated at $z=z_{uv}$ remain finite as $z_{uv} \to \infty$, while the coefficients of others may not be fixed a priori.  The latter sort of terms are called `finite counterterms' in the literature, and the freedom to choose their coefficients reflects the freedom to choose a renormalization scheme.  Physical observables must of course be scheme-independent, but the detailed connection between `bare' model parameters and physical observables is in general scheme-dependent.   The observable we will be calculating, Eq.~\eqref{eq:MagicRatio}, is a pure number, and by construction cannot depend on any model parameters in any theory consistently implementing the expectations of chiral perturbation theory.  Thus it also cannot depend on any of the boundary counterterms in $S_{b}$, and as a result in this paper we will not need to work with $S_{b}$ in an explicit form.  It is nonetheless important to keep in mind that in principle $S_{b} \neq 0$, and its detailed form should be expected to affect the mapping between the model parameters and field theory observables.

Ultimately we seek to compute the electromagnetic form factors of the baryon, which are evaluated as matrix elements of vector and axial isospin currents.  The currents can be computed using the standard ideas of gauge-gravity duality.  The gauge theory lives on the UV boundary ($z=z_{\mathrm{uv}}$ in our case) of the gravity theory.   Each gauge-invariant operator in the gauge theory is associated with a dynamical field in the gravity theory, with the boundary value of the bulk field acting as a source for the operator.  The partition function of the field theory is equated to the partition function of the gravity theory with all bulk fields evaluated on-shell.   In particular, since the source for a conserved current in a field theory is a gauge field, the bulk fields associated with conserved currents are gauge fields, which are precisely the $\A$ fields in the action above.  As shown in {\it e.g.} Ref.~\cite{Hashimoto:2008zw} in the Sakai-Sugimoto model the isovector $J_{V}$, axial isovector $J_{A}$, isoscalar $\hat{J}_{S}$ and axial isoscalar $\hat{J}_{A}$ currents are
\begin{align}
\label{eq:CurrentDefinitions}
J^{a}_{V, \mu} &= - \kappa \left[k(z) F^{a}_{\mu z} \right]^{z=z_{\mathrm{uv}}}_{z=-z_{\mathrm{uv}}} , \\
J^{a}_{A, \mu} &= - \kappa \left[\frac{2}{\pi} \tan^{-1}(z) k(z) F^{a}_{\mu z} \right]^{z=z_{\mathrm{uv}}}_{z=-z_{\mathrm{uv}}} , \\
J_{S, \mu} &= - \kappa \left[ k(z) \hat{F}_{\mu z} \right]^{z=z_{\mathrm{uv}}}_{z=-z_{\mathrm{uv}}} , \\
J_{A, \mu} &= - \kappa \left[\frac{2}{\pi} \tan^{-1}(z) k(z) \hat{F}_{\mu z} \right]^{z=z_{\mathrm{uv}}}_{z=-z_{\mathrm{uv}}} ,
\end{align}
where the bulk fields are evaluated on solutions to the bulk equations of motion, and we split $\A$ (and hence $\F$) into $SU(2)$ and $U(1)$ pieces as $\A = A+\frac{1}{2} \mathbf{1}_{2} \hat{A}$.

%%%%%%%%%%%%%%%
\subsection{Baryons as solitons}
\label{sec:BaryonAnsatz}
%%%%%%%%%%%%%%%

Baryons are encoded in soliton solutions of the model. In the D-brane construction, a D4-brane wrapped on $S^4$ provides a baryon vertex in the bulk\cite{Witten:1979kh},  and appears as instanton-like solitonic configurations in the gauge theory on flavor D8-branes\cite{Sakai:2004cn,Sakai:2005yt,Hata:2007mb,Hong:2007ay,Hong:2007kx,Hashimoto:2005qh}.  In the effective 5D YM+CS theory, the baryon solitons carry unit instanton number on $x_{1},x_{2}, x_{3}, z$~\footnote{ For some recent discussions of 5D instantons in supersymmetric YM+CS theories in flat space, see Refs.~\cite{Collie:2008vc,Kim:2008kn}.}.  We start from the static solutions for simplicity. Collective coordinate quantization of the soliton, which is necessary to pick out the physics of protons from the physics of solitons, will be considered later.

There are a number of symmetries we can take advantage of to aid in finding the solutions.  On general grounds, we expect that the ground-state baryons in the exact isospin limit will be spherically symmetric.  On the gravity side, the action has an $SO(3)$ spatial symmetry, and so it is  natural to expect that the minimum-energy static soliton will be $SO(3)$ symmetric, provided that the boundary conditions also have this symmetry.  In the usual approach to baryons in the Sakai-Sugimoto model, which relies on the flat-space instanton approximation, one essentially seeks an $SO(4)$ symmetric configuration, since there are some arguments that an approximate $SO(4)$ symmetry appears at large $\lambda$.  Here we want to avoid relying on the flat-space instanton approach, and do not ask for $SO(4)$ symmetry.  In any case, the $SO(3)$-symmetric ansatz is the only self-consistent  ansatz one can impose {\it a priori} (without making approximations), since once we  add the point at spatial infinity the topology of our manifold is $S^{3} \times I$, not $S^{4}$.  On top of this any putative $SO(4)$ symmetry of the geometry is broken by the warp functions $h(z), k(z)$.  

The correct procedure for imposing spacetime symmetries on gauge fields was elucidated in the work of Forgacs and Manton\cite{Forgacs:1979zs}, following Witten\cite{Witten:1976ck}.  In the context of holographic baryons, this procedure was first applied by Panico, Pomarol and Wulzer\cite{Pomarol:2007kr,Pomarol:2008aa,Panico:2008it}, whose notation we will follow.  In particular, it turns out that consistently imposing an $SO(3)_{s}$ rotation symmetry on a $U(2)$ gauge field $\A_{M}$ necessitates demanding that $\A_{M}$ be invariant under the diagonal combination $SO(3)_{s} + SO(3)_{g}$ transformations, where the second factor is the $SO(3)$ subgroup of the gauge group $U(2)$. This procedure produces a self-consistent $SO(3)$ symmetric ansatz for $\A_{M}$ that can be plugged back into the action.

For time-independent configurations, the YM-CS action has a $Z_{2}$ symmetry which implies that we can set $A_{0}$ and the spatial components of $\hat{A}$ to zero\cite{Hata:2007mb,Pomarol:2008aa}.  The $SO(3)$-symmetric time-independent ansatz is then given by\cite{Witten:1976ck,Forgacs:1979zs,Pomarol:2007kr,Pomarol:2008aa,Panico:2008it}
\begin{align}
\label{eq:StaticAnsatz}
A_{j}^{a} &= \frac{\phi_{2}+1}{r^{2}} \epsilon_{j a k} x_{k} + \frac{\phi_{1}}{r^{3}} [\delta_{ja} r^{2} - x_{j} x_{a}]+A_{r} \frac{x_{j}x_{a}}{r^{2}}, \nonumber \\
A_{z}^{a} &= A_{z} \frac{x^{a}}{r},  \quad \hA_{0} = s.
\end{align}
where $\phi_{1},\phi_{2},A_{z},A_{r},s$ are all functions of $r$ and $z$; $i,j = 1,2,3$ run over the spatial indices, and $a=1,2,3$ is an isospin index.  The fields $\phi_{1},\phi_{2}$ can be naturally packaged into a complex scalar field $\phi = \phi_{1}+i\phi_{2}$ which has unit charge under the Abelian gauge field with components $A_{z},A_{r}$, so that $D_{\mu} \phi = \partial_{\mu}\phi +  iA_{\mu} \phi$, where now $\mu = z,r$.  The reason for the presence of the 2D Abelian gauge field $A_{\mu}$ in the  ansatz can be traced to an unbroken Abelian subgroup of original $SU(2)$ gauge symmetry in the symmetry-reduced ansatz.

Plugging in the ansatz, the mass-energy of the reduced system can be written as 
\begin{align}
\label{eq:StaticSolitonEnergy}
M=M_{YM}+M_{CS}+M_{b} ,
\end{align}
where $\int dt M = S_{5D}$, $\int dt M_{b} = S_{b}$,
\begin{align}
M_{YM} = 16 \pi \kappa \int_{0}^{\infty}dr \int_{-z_\mathrm{uv}}^{z_\mathrm{uv}} dz \,  &\left[  h(z) |D_{r} \phi|^{2} + k(z) |D_{z} \phi|^{2}  +\frac{1}{4}r^{2} k(z) F_{\mu\nu}^{2} \right.  \\
&\left. +\frac{1}{2r^{2}} h(z) (1-|\phi|^{2})^{2} -\frac{1}{2}r^{2} \left(h(z) (\partial_{r} s)^{2} + k(z) (\partial_{z} s)^{2}\right) \right],
\end{align} 
and
\begin{align}
M_{CS} =16\pi \kappa \gamma \int_{0}^{\infty}{dr \int_{-z_\mathrm{uv}}^{z_\mathrm{uv}}  dz \, s\,\epsilon^{\mu \nu}\left[ \partial_{\mu}(-i\phi^{*}D_{\nu}\phi + h.c) +F_{\mu \nu}\right]} ,
\end{align}
with $\gamma = N_{c}/(16\pi^{2} \kappa) = 27\pi/(2\lambda)$. The expression for the topological charge in terms of the reduced ansatz is 
\begin{align}
\label{eq:TopologicalCharge2D}
Q = \frac{1}{4\pi} \int{dr dz \, \left(  \epsilon^{\mu \nu}\partial_{\mu}\left[ - i \phi^{*}D_{\nu}\phi + h.c.\right] + \epsilon^{\mu \nu} F_{\mu \nu} \right)}.
\end{align}
Since by definition $S_{b}$ is a pure boundary term, it does not contribute to the bulk equations of motion.  We will also shortly see that with our choice of boundary conditions $S_{b}$ manifestly cannot contribute to $Q$.  The equations of motion that follow from extremizing $M_{YM}+M_{CS}$ are
\begin{align}
\label{eq:2DEoMZcoord}
0 &= D_{r} \left(h(z) D_{r} \phi \right)+D_{z} \left(k(z) D_{z} \phi \right) + \frac{h(z)}{r^{2}} \phi(1-|\phi|^{2}) -i \gamma \epsilon^{\mu \nu} \partial_{\mu} s  D_{\nu} \phi , \\
0 &=\partial_{r} \left(r^{2} k(z) F_{r z}\right) - k(z) \left(i \phi^{\ast} D_{z}\phi + h.c. \right) - \gamma \epsilon^{r z} \partial_{r}s(1-|\phi|^{2}) , \\
0 &=\partial_{z} \left(r^{2} k(z) F_{z r}\right) - h(z) \left(i \phi^{\ast} D_{r}\phi + h.c. \right) - \gamma \epsilon^{z r} \partial_{z}s(1-|\phi|^{2}) , \\
0 &=\partial_{r} \left( h(z) r^{2} \partial_{r} s \right) +\partial_{z} \left( k(z) r^{2} \partial_{z} s \right)  + \gamma \epsilon^{\mu \nu} \left[ \partial_{\mu}(-i\phi^{\ast}D_{\nu}\phi + h.c) +F_{\mu \nu}\right].
\end{align}

The equations of motion are a set of coupled second-order non-linear PDEs, and a general analytic solution of them seems out of reach. In general one would have to construct the solutions numerically.  Fortunately, given our goal of investigating the long-distance properties of the baryon, we only need to know the behavior of the solution at large distance $r$.  For this we only need to solve the equations of motion to the first few orders in a $1/r$ expansion, and this can be done analytically.  Implicit in our analysis will be the assumption that there exists a global unique and well-behaved solution to the boundary value problem with $Q=1$.
 
We wish to choose boundary conditions which will give finite-energy solutions with charge $Q=1$.  The integrand in the Eq.~\eqref{eq:TopologicalCharge2D} is a total derivative, and $Q$ can be written as 
\begin{align}
Q &= \frac{1}{2\pi} \int_{0}^{R}{dr \, \left[ \frac{1}{2}\left(-i \phi D_{r} \phi^{*} +i \phi^{*} D_{r} \phi \right) + A_{r}\right]\Big|^{z=+z_{\mathrm{uv}}}_{z=-z_{\mathrm{uv}}}}   \\
&-\frac{1}{2\pi}  \int_{-z_{\mathrm{uv}}}^{z_{\mathrm{uv}}}{dz \, \left[ \frac{1}{2}\left(-i \phi D_{z} \phi^{*} +i \phi^{*} D_{z} \phi \right) + A_{z}\right]\Big|^{r=R}_{r=0}},
\end{align}
with $R\to \infty$. To find the large-$r$ solutions, we find it convenient to work in the 2D Lorentz gauge $\partial_{\mu}A^{\mu}=0$, and choose the boundary conditions for the fields $\phi, A_{\mu},s$ in such a way that the only non-trivial contribution to $Q$ comes from the boundary at $r=0$.   These boundary conditions are summarized in Table~\ref{table:StaticBoundaryConditions}.  With this choice of boundary conditions $S_{b}$ certainly cannot contribute to $Q$.  The 5D $SU(2)$ gauge fields are singular at $r=0$ in this gauge, but of course gauge-invariants like the field strength remain smooth at the origin.

%%%%%%%%%%%%%%%%%%%%
\begin{table}[htdp]
\begin{center}\begin{tabular}{|c|c|c|} \hline $r \to \infty$ & $r = 0$ & $z=\pm z_{\mathrm{uv}}$  \\
 \hline 
 $\phi_{1} = 0$  & $\phi_{1}  =  \sin \left(\frac{2 \pi z}{L} \right)$ & $\phi_{1} =0$  \\ 
 $\phi_{2} =  -1$  & $\phi_{2} = - \cos \left(\frac{2 \pi z}{L} \right)$ & $\phi_{2} = -1$    \\
 $A_{z} =0$ & $A_{z} = \frac{\pi}{L}$ & $\partial_{z}A_{z} = 0$  \\
 $\partial_{r}A_{r}=0$ & $\partial_{r}A_{r} = 0$ & $A_{r} = 0$  \\
 $s=0$ & $s=0$ & $s=0$ \\
 \hline
 \end{tabular}
\caption{Boundary conditions for $\phi = \phi_{1}+i\phi_{2}$, $A_{\mu}$ and $s$ in the 2D Lorentz gauge $\partial_{\mu} A^{\mu} = 0$.  Here $L = 2z_{\mathrm{uv}}$ is the extent of the holographic direction with the UV cutoff $z_{\mathrm{uv}}$. }
\label{table:StaticBoundaryConditions}
\end{center}
\end{table}
%%%%%%%%%%%%%%%%%%%%

%%%%%%%%%%%%%%%%
\section{Large $r$ properties of baryons}
\label{sec:LargeR}
%%%%%%%%%%%%%%%%
We are interested in the solutions at large $r$.  If $r$ is much larger than the characteristic physical scale of the model, $\MKK^{-1}$, the equations of motion should linearize.  This motivates looking for asymptotic solutions to the equations of motion as a Taylor series in $1/r$, with coefficient functions that depend on $z$.  If the coefficient functions of the Taylor expansion can be solved for order by order, the linearization of the equations of motion at large $r$ will have been demonstrated self-consistently.  It is important to note the physics behind these statements.  Seeking solutions to the equations of motion for a soliton which are Taylor series in $1/r$ at large $r$ is tantamount to asking that the profile function of the soliton have a long-range pion tail.  Showing that the coefficient functions of the Taylor expansion are non-trivial and can be solved for order by order then shows self-consistently that the profile function indeed has a long-range pion tail, and hence implies that the baryon is surrounded by a pion cloud.

%%%%%%%%%%%%%%%
\subsection{Static solutions}
\label{sec:StaticSolitons}
%%%%%%%%%%%%%%%

To set up the expansion we write $\phi_{1},\phi_{2}, A_{z},A_{r},s$ as
\begin{align}
\phi_{1}(r,z) &= \sum_{n=1}^{\infty} \phi^{(n)}_{1}(z) \frac{1}{r^{n}}, \;\; 
\phi_{2}(r,z) = -1 + \sum_{n=1}^{\infty} \phi^{(n)}_{2}(z) \frac{1}{r^{n}}, \\
A_{z}(r,z) &= \sum_{n=1}^{\infty} A^{(n)}_{z}(z) \frac{1}{r^{n}}, \;\;
A_{r}(r,z) = \sum_{n=1}^{\infty} A^{(n)}_{r}(z) \frac{1}{r^{n}},  \;\; 
s(r,z) = \sum_{n=1}^{\infty} s^{(n)}_{r}(z) \frac{1}{r^{n}}.
\end{align}
This ansatz is chosen to automatically satisfy the boundary conditions at $r=\infty$.  The equations of motion then reduce to a set of coupled ODEs for the coefficient functions  $\phi^{(n)}_{1}(z),\phi^{(n)}_{2}(z),A^{(n)}_{z}(z),A^{(n)}_{r}(z),s^{(n)}(z)$.  It is then possible to solve for the coefficient functions order by order in $1/r$ using the boundary conditions at $z=\pm z_{\mathrm{uv}}$.  Counting boundary conditions we see that we have enough data to fix all of the solutions order by order in terms of a single undetermined constant of integration, which we call $\beta$ below.  

The role of $\beta$ is to parametrize the dependence of the large-$r$ solution on the rest of the solution.  To see this, note that the physical boundary value problem that determines the full soliton field configuration depends on two parameters, $z_{\mathrm{uv}}$ and $\gamma$~\footnote{Assuming that the topological charge is fixed to $Q=1$.}. If one makes the standard physical assumption that there exists a unique well-behaved solution to the boundary value problem with a given $z_{\mathrm{uv}}$ and $\gamma$, then $\beta$ will be fixed by matching to full solution, and will be a function of $z_{\mathrm{uv}}$ and $\gamma$.

Solving the equations of motion at large-$r$ order-by-order in power series in $1/r$, we find that the first few terms in the expansion are
{\allowdisplaybreaks
 \begin{align}
\label{eq:StaticSolutions}
\phi_{1} &= \frac{\beta (z -  \frac{z_{\mathrm{uv}}\tan^{-1}(z)}{\tan^{-1}(z_{\mathrm{uv}})})}{r^{2}} - \\ 
&\frac{\beta  \left(-3 \left(-1+z^2\right) z_{\mathrm{uv}} \tan^{-1}z+z \left(\left(-3+z^2+2 z_{\mathrm{uv}}^2\right) \tan^{-1}z_{\mathrm{uv}}+3 z_{\mathrm{uv}} \log \left[\frac{1+z^2}{1+\zuv^{2}}\right]\right)\right)}{\tan^{-1}[z_{\mathrm{uv}}] r^4} \nnb ,\\
\phi_{2} &= -1+\frac{\frac{1}{2} \left(z^2+z_{\mathrm{uv}}^2\right) \beta ^2-\frac{z z_{\mathrm{uv}} \beta ^2 \tan^{-1}[z]}{\tan^{-1}[z_{\mathrm{uv}}]}}{r^4} , \\
A_{z} &= \frac{\beta }{r^2}+\frac{\beta \left(6 z z_{\mathrm{uv}} \tan^{-1}[z]+\left(3-3 z^2-2 z_{\mathrm{uv}}^2\right) \tan^{-1}[z_{\mathrm{uv}}]-3 z_{\mathrm{uv}} \left(1+\log \left[\frac{1+z^2}{1+\zuv^{2}}\right]\right)\right)}{\tan^{-1}[z_{\mathrm{uv}}] r^4} , \\
A_{r} &=\frac{-2 z \beta +\frac{2 z_{\mathrm{uv}} \beta  \tan^{-1}[z]}{\tan^{-1}[z_{\mathrm{uv}}]}}{r^3}+ \\
&\frac{4 \beta  \left(-3 \left(-1+z^2\right) z_{\mathrm{uv}} \tan^{-1}[z]+z \left(\left(-3+z^2+2 z_{\mathrm{uv}}^2\right) \tan^{-1}[z_{\mathrm{uv}}]+3 z_{\mathrm{uv}} \log \left[\frac{1+z^2}{1+\zuv^{2}}\right]\right)\right)}{\tan^{-1}[z_{\mathrm{uv}}] r^5} , \nnb \\
s &= \frac{\beta ^3 \gamma z_\mathrm{uv}^3 ( \tan ^{-1}(z)^4-6 \tan ^{-1}(z)^2 \tan ^{-1}(z_{\mathrm{uv}})^2+5 \tan ^{-1}(z) \tan ^{-1}(z_{\mathrm{uv}})^3)}{2 \tan ^{-1}(z_{\mathrm{uv}})^3 r^{9}} .
\end{align}}

It is crucial to work with a finite cutoff $\zuv$ to obtain these solutions.  If one tries to take $\zuv \to \infty$ when looking for the solutions, only the trivial solution to the boundary-value problem will be found.  To see this from the solutions above, consider for instance $\phi_{1}$. At large $r$,
 \begin{align}
 \lim_{r\to \infty} r^{2}\phi_{1} =  \beta \left(z -  \frac{z_{\mathrm{uv}}\tan^{-1}(z)}{\tan^{-1}(z_{\mathrm{uv}})} \right)
 \equiv \Phi_{z}(z;\zuv)  
\end{align}
In the large $\zuv$ limit, $\Phi \to - \frac{2\beta \zuv}{\pi} \tan(z)$, which solves the differential equation for $\phi_{1}$.  However, the only way to satisfy the boundary condition $\phi_{1} \to 0$ as $z\to \infty$ is to set $\beta = 0$. So if one tries to work without a cutoff on $z$, as has thus far been the approach in the literature, the only solutions for the boundary value problem which are power series in $1/r$ that can be obtained are the trivial ones.  One then cannot capture all of the long-distance physics.  With a cutoff, however, one obtains non-trivial solutions to the boundary value problem order by order in a $1/r$ expansion, and the model has a chance of capturing the physics of interest.  

With the above solutions in hand, we can evaluate the asymptotic on-shell action density for static solutions at large $r$, with the result that 
\begin{align}
(S - S_{b}) |_{r\to\infty}  &= \int dt dz \left\{\frac{3 (z_{\mathrm{uv}}\beta)^{2}}{(1+z^{2}) \tan^{-1}(z_{\mathrm{uv}}) r^{4}} +\mathcal{O}(1/r^{8}) \right\} 
= \int dt \left\{\frac{6 (z_{\mathrm{uv}}\beta)^{2}}{ \tan^{-1}(z_{\mathrm{uv}})} +\mathcal{O}(1/r^{8})   \right\} \\
& \to  \int dt \left\{\frac{12 (z_{\mathrm{uv}}\beta)^{2}}{\pi r^{4}} +\mathcal{O}(1/r^{8}) \right\}  ,
\end{align}
where in the last line we took $z_{\mathrm{uv}}\gg 1$.  There are then two possibilities for the behavior of the on-shell action as $\zuv \to \infty$: either $\beta$ behaves as $\beta \sim 1/\zuv$ at large $\zuv$ in the full $Q=1$ finite-energy solution, or it does not.  If $\beta$ behaves as $\beta  \sim 1/z_{\mathrm{uv}}$, then the on-shell action will have no large $z_{\mathrm{uv}}$ divergences at large $r$, and $S_{b}$ can only contain finite counterterms.  Otherwise, the on-shell action without $S_{b}$ may be divergent at large $\zuv$, and then $S_{b}$ would also contain terms that diverge at large $\zuv$ in such a way that the full on-shell action is finite.   A very interesting task we leave to future work is to work out the behavior of the full solution and understand which of these options is the relevant one  by solving the full PDE and examining the asymptotic behavior of the solution.   Fortunately for the current analysis, the evaluation of the form-factor ratio is independent of these interesting and subtle issues.

%%%%%%%%%%%%%%%%%%%%%%%
\subsection{Rotating solutions}
\label{sec:RotatingSolitons}
%%%%%%%%%%%%%%%%%%%%%%%%

To extract information about states with the quantum numbers of the proton, we must follow the procedure of collective-coordinate quantization\cite{Adkins:1983ya}.   This is because in general, single-particle states associated with solitons are described as quantized time-dependent fluctuations around the static soliton solution in the zero mode directions.
To perform collective coordinate quantization, one allows the soliton to slowly rotate in the zero-mode directions.  Then the gauge field components $A_{0}$ and $\hat{A}_{i}, \hat{A}_{z}$ must be turned on so as to parametrize the  collective motion of the rotating soliton.   In the limit of very slow rotation (ultimately justified by the large $N_{c}$ limit, which makes the moment of inertia of the soliton large), the shape of the soliton cannot be affected by the rotation, so we expect the soliton to keep its $SO(3)$ rotational symmetry.  This $SO(3)$ symmetry can then be used to constrain the form of $A_{0}$ and $\hat{A}_{i}, \hat{A}_{z}$.  The general $SO(3)$ symmetric ansatz appropriate to a soliton rotating with a constant angular velocity $\vec{k}$ is given by \cite{Panico:2008it}
\begin{align}
\label{eq:TimeDependentAnsatz}
A^{a}_{0} &= k_{b}\left[\chi_{1}\epsilon^{abc}\hat{x}_{c} +\chi_{2} (\hat{x}^{a} \hat{x}^{b} - \delta^{ab}) \right] + v (\vec{k} \cdot \hat{x}) \hat{x}^{a} ,\\
\hat{A}_{i} &=\frac{\rho}{r} (k_{i} - (\vec{k}\cdot \hat{x})x_{i}) +B_{r}  (\vec{k}\cdot \hat{x}) \hat{x}_{i} + Q \epsilon_{i b c} k^{b}\hat{x}^{c} ,\\
\hat{A}_{z} &= B_{z}  (\vec{k}\cdot \hat{x}) .
\end{align}
The way the new fields $\chi_{1,2}, \rho, B_{r,z}, Q, v$ come into the symmetry-reduced action is constrained by the residual gauge symmetry.  The field $\chi = \chi_{1}+i\chi_{2}$ transforms the same way as $\phi$, and couples to the gauge field $A_{\mu}$. The time-dependent ansatz has a new residual $U(1)$ gauge symmetry, associated with chiral $U(1)_{L,R}$ transformations of the form $\hat{g}_{R} = g$, $\hat{g}_{L} = g^{\dagger}$ with
\begin{align}
g = \exp\left(i \xi(r,z)  (\vec{k}\cdot \hat{x}) \right).
\end{align}
Under this transformation only $B_{\mu}$ and $\rho$ transform non-trivially, with transformation rules
\begin{align}
B_{\mu} \to B_{\mu} + \partial_{\mu} \xi, \;\; \rho \to \rho  +\xi.
\end{align}

Evaluated on the $SO(3)$ symmetric time-dependent ansatz, the YM-CS Lagrangian becomes
\begin{align}
\label{eq:RotatingSoliton}
\mathcal{L} = -M + \frac{\Lambda}{2} k_{a} k^{a}
\end{align}
to first order in $\vec{k}$, where $M$ is the mass of the static configuration and was given in Eq.~\eqref{eq:StaticSolitonEnergy}, and the moment of inertia $\Lambda$ can be written as
\begin{align}\label{eq:moment_of_inertia}
\Lambda =&\frac{16\pi \kappa}{3}\int_0^\infty dr\int^{-z_\mathrm{uv}}_{z_\mathrm{uv}}dz\,
\Big[
-h(z) (D_r\rho)^2 - k(z) (D_z\rho)^2
-r^2 h(z) (\partial_r Q)^2 - r^2 k(z) (\partial_z Q)^2
\nonumber \\
& -2h(z) Q^2
-\frac{r^2}{2} h(z) B_{\mu\nu}^2
+r^2 h(z) |D_r\chi|^2 + r^2 k(z) |D_z\chi|^2 
+\frac{r^2}{2} h(z)(\partial_r v)^2 + \frac{r^2}{2} k(z)(\partial_z v)^2
\nonumber \\
&+ h(z) (|\chi|^2+v^2)(1+|\phi|^2)
-4v\phi_p\chi_p
+\gamma
\Big(
-2\epsilon^{\mu\nu}D_{\mu}\rho \, \chi_p (D_{\nu}\phi)_p
\nonumber \\
&+2\epsilon^{\mu\nu}\partial_{\mu}(r Q) \chi_{p}\epsilon^{pq}(D_{\nu}\phi)_q
-v(\epsilon^{\mu\nu}B_{\mu\nu}(|\phi|^2 -1)/2
+r Q\epsilon^{\mu\nu}A_{\mu\nu}
+2rQ\epsilon^{\mu\nu} D_\mu \rho\partial_\nu s
\Big] \Big],
\end{align}
where $p,q=1,2$.  Note that $\Lambda \sim N_{c}$.  The form of Eq.~\eqref{eq:RotatingSoliton} is that of a Lagrangian for a rigid rotor of mass $M$ and moment of inertia $\Lambda$, with collective coordinates $k_{a}, a=0,1,2,3$.   The equations of motion that follow from this Lagrangian are
{\allowdisplaybreaks
\begin{align}\label{eq:collective_eom}
& \frac{1}{r^2}\partial_r (r^2 \partial_r v) + \frac{1}{h(z)} \partial_z( k(z) \partial_z v) - \frac{2}{r^2} ( v(1+|\phi|^2)- \chi \phi^\dagger -\phi \chi^\dagger )
+ \frac{\gamma}{r^2 h(z)}[(|\phi|^2-1) B_{rz} + 2rQF_{rz}]=0, \nonumber \\
& \frac{1}{r^2} D_r (r^2 D_r \chi) + \frac{1}{h(z)} D_z (k(z) D_z \chi) + \frac{1}{r^2} (2v\phi-(1+|\phi|^2)\chi)
- \frac{\gamma }{r^2 h(z)} \epsilon^{\mu\nu}(D_\mu \phi (i \partial_\nu Q +D_\nu \rho)=0, \nonumber \\
& \frac{1}{r^2}\partial_r (r^2 \partial_r Q) + \frac{1}{h(z)} \partial_z (k(z)\partial_z Q) - \frac{2}{r^2} Q \nonumber \\
& \qquad- \frac{\gamma }{ 2r h(z)} \epsilon^{\mu\nu}[ (i D_\mu \phi(D_\nu \chi)^\dagger + h.c. )+ F_{\mu\nu} (2v-\chi\phi^\dagger -\phi\chi^\dagger)/2 -2 D_\mu\rho \, \partial_\nu s]=0, \nonumber \\
& \partial_r (D_r \rho) + \frac{1}{h(z)} \partial_z (k(z) D_z \rho)
-\frac{\gamma }{2 h(z)} \epsilon^{\mu\nu} [(D_\mu \phi (D_\nu \chi)^\dagger + h.c.) + i F_{\mu\nu}(\phi\chi^\dagger-\chi\phi^\dagger)/2+2\partial_\mu(rQ)\partial_\nu s ]=0, \nonumber \\
& \frac{1}{h(z)}\partial_z(k(z) B_{zr}) + \frac{2}{r^2}D_r \rho
+\frac{\gamma}{r^2 h(z)} [ ((\chi-v\phi)(D_z \phi)^\dagger + h.c.) + (1-|\phi|^2) \partial_z v - 2r Q\partial_z s]=0, \nonumber \\
& \frac{1}{r^2}\partial_r(r^2 B_{rz}) + \frac{2}{r^2}D_z \rho
-\frac{\gamma}{r^2 k(z)} [ ((\chi-v\phi)(D_r \phi)^\dagger + h.c.) + (1-|\phi|^2) \partial_r v - 2r Q\partial_r s]=0.
\end{align} }
The boundary conditions for the new fields are given in Table \ref{tab:collective_bc}.
%%%%%%%%%%%%%%%%%%%%
\begin{table}[htdp]\label{tab:collective_bc}
\begin{center}\begin{tabular}{|c|c|c|} \hline $r \to 0$ & $r = \infty$ & $z=\pm z_{\mathrm{uv}}$  \\
 \hline 
 $\chi_{1} = -\sin \left(\frac{2 \pi z}{L} \right)$  & $\chi_{1}  =  0$ & $\chi_{1} =0$  \\ 
 $\chi_{2} =   \cos \left(\frac{2 \pi z}{L} \right)$  & $\chi_{2} = 1$ & $\chi_{2} = -1$    \\
 $B_{z} =0$ & $B_{z} = 0$ & $\partial_{z}B_{z} = 0$  \\
 $\partial_{r}B_{r}=0$ & $\partial_{r}B_{r} = 0$ & $B_{r} = 0$  \\
 $\rho=0$ & $\rho=0$ & $\rho = 0$ \\
 $v=-1$ & $v=-1$ & $v=-1$ \\
 $Q=0$ & $Q=0$ & $Q=0$ \\
 \hline
 \end{tabular}
\caption{Boundary conditions for $\chi = \chi_{1}+i\chi_{2}, B_{\mu}, Q, v$ and $\rho$ in the 2D Lorentz gauge $\partial_{\mu} B^{\mu} = 0$.   }
\end{center}
\label{table:StaticBoundaryConditions}
\end{table}
%%%%%%%%%%%%%%%%%%%%

Using the same approach as in the static case, it is straightforward to solve the equations of motion \eqref{eq:collective_eom} to the first few orders in a $1/r$ expansion.  The solution depends on the matching parameter $\beta$ already seen in the static solutions, and on an additional matching parameter $\tilde{\beta}$.  The results are
{\allowdisplaybreaks
\begin{align}
\label{eq:TimeDependentSolutions}
\chi_{1} &= \frac{-z \beta +\frac{z_{\mathrm{uv}} \beta  \tan^{-1}(z)}{\tan^{-1}(z_{\mathrm{uv}})}}{r^2}+ \\
&\frac{\beta  \left(-3 \left(-1+z^2\right) z_{\mathrm{uv}} \tan^{-1}(z)+z \left(-3+z^2+2 z_{\mathrm{uv}}^2\right) \tan^{-1}(z_{\mathrm{uv}})+3 z z_{\mathrm{uv}} \log\left[\frac{1+z^2}{1+z_{\mathrm{uv}}^2}\right]\right)}{\tan^{-1}(z_{\mathrm{uv}}) r^4} \nnb\\
\chi_{2} &= 1+\frac{-\frac{1}{2} \left(z^2+z_{\mathrm{uv}}^2\right) \beta ^2+\frac{z z_{\mathrm{uv}} \beta ^2 \tan^{-1}(z)}{\tan^{-1}(z_{\mathrm{uv}})}}{r^4} \\
B_{r} &= \frac{-2 \tilde{\beta} z+\frac{2 \tilde{\beta} z_{\mathrm{uv}} \tan^{-1}(z)}{\tan^{-1}(z_{\mathrm{uv}})}}{r^3}+  \\
&\frac{4 \tilde{\beta} \left( \frac{-3 \left(-1+z^2\right) z_{\mathrm{uv}}\tan^{-1}(z)}{\tan^{-1}(z_{\mathrm{uv}})}+z \left(-3+z^2+2 z_{\mathrm{uv}}^2\right) +\frac{3 z z_{\mathrm{uv}} }{\tan^{-1}(z_{\mathrm{uv}})}\log\left[\frac{1+z^2}{1+z_{\mathrm{uv}}^2}\right]\right)}{r^5}  \nnb \\
B_{z} &= \frac{\tilde{\beta}}{r^2}+\frac{\tilde{\beta} \left( \frac{6 z z_{\mathrm{uv}}\tan^{-1}(z)}{\tan^{-1}(z_{\mathrm{uv}})}+\left(3-3 z^2-2 z_{\mathrm{uv}}^2\right) - \frac{3z_{\mathrm{uv}} }{\tan^{-1}(z_{\mathrm{uv}})}\left(1+\log \left[\frac{1+z^2}{1+\zuv^{2}}\right]\right)\right)}{ r^4}  \nnb\\
v &= -1 + \mathcal{O}(1/r^{6})  \\
Q &= -\frac{z_{\mathrm{uv}}^3 \beta ^3 \gamma  \left(\tan^{-1}(z)^4-6 \tan^{-1}(z)^2 \tan^{-1}(z_{\mathrm{uv}})^2+5 \tan^{-1}(z_{\mathrm{uv}})^4\right)}{2 \tan^{-1}(z_{\mathrm{uv}})^3 r^8} , \\
\rho &= \frac{\tilde{\beta} \left(z-\frac{z_{\mathrm{uv}} \tan^{-1}(z)}{\tan^{-1}(z_{\mathrm{uv}})}\right)}{r^2}+ \\
&- \frac{\tilde{\beta} \left(-3 \left(-1+z^2\right) z_{\mathrm{uv}} \tan^{-1}(z)+z \left(-3+z^2+2 z_{\mathrm{uv}}^2\right) \tan^{-1}(z_{\mathrm{uv}})+3 z z_{\mathrm{uv}} \log\left[\frac{1+z^2}{1+z_{\mathrm{uv}}^2}\right]\right)}{\tan^{-1}(z_{\mathrm{uv}}) r^4} . \nnb
\end{align}}

Evaluating the asymptotic on-shell moment of inertia at large $r$, we obtain
\begin{align}
\Lambda |_{r\to\infty}&=
\frac{16\pi\kappa}{3} \int dz  \left\{
-\frac{(z_\mathrm{uv}\beta)^2}{k(z)(\tan^{-1}z_\mathrm{uv})^2 r^2}
+\frac{3 (z_\mathrm{uv}\tilde{\beta})^2}{k(z) (\tan^{-1}z_\mathrm{uv})^2 r^4}
+\mathcal{O}(1/r^6)
\right\} \nonumber \\
&=\frac{16\pi\kappa}{3}  \left\{
-\frac{2 (z_\mathrm{uv} \beta)^2}{\tan^{-1}(z_\mathrm{uv}) r^2}
+\frac{6 (z_\mathrm{uv}\tilde{\beta})^2}{\tan^{-1}(z_\mathrm{uv}) r^4}
+\mathcal{O}(1/r^6)
\right\} \nonumber \\
& \to \frac{16\pi\kappa}{3} \left\{
-\frac{4 (z_\mathrm{uv} \beta)^2}{\pi r^2}
+\frac{12 (z_\mathrm{uv}\tilde{\beta})^2}{\pi r^4}
+\mathcal{O}(1/r^6)
\right\}.
\end{align}
We observe that for the rotating solution to have finite energy, either $\tilde{\beta}$ must go to zero at least as fast as $\sim 1/z_\mathrm{uv}$, or the $\tilde{\beta}$ contribution must be cancelled by a contribution from the boundary action $S_{b}$.  Fortunately, as will be seen shortly $\tilde{\beta}$ does not contribute to the leading large-$r$ behavior of the form factors, and so we can defer an investigation of the behavior of $\tilde{\beta}$ to future work.

%%%%%%%%%%%%%%%%%%%%%%%%
\subsection{Evaluation of the form factors}
%%%%%%%%%%%%%%%%%%%%%%%%

In Ref.~\cite{Cherman:2009gb} it was argued that \XPT\ constrains the Fourier transforms of the electromagnetic form factors of the proton to take the large-$r$ forms
\begin{align}
\label{eq:FF_GSE}
G_{I=0}^{E} &\to \frac{3^3}{2^9\pi^5} \frac{1}{f_{\pi}^3}\left(\frac{g_A}{f_{\pi}}\right)^3 \frac{1}{r^9} ,\\
\label{eq:FF_GSM}
G_{I=0}^{M} &\to \frac{3\Delta}{2^9  \pi^5}\frac{1}{f_{\pi}^{3}} \left(\frac{g_A}{f_{\pi}}\right)^3 \frac{1}{r^7} ,\\
\label{eq:FF_GVE}
G_{I=1}^{E} &\to \frac{\Delta}{2^4  \pi^2} \left(\frac{g_A}{f_{\pi}}\right)^2 \frac{1}{r^4} , \\
\label{eq:FF_GVM}
G_{I=1}^{M} &\to \frac{1}{2^5 \pi^2} \left(\frac{g_A}{f_{\pi}}\right)^2 \frac{1}{r^4} .
\end{align}
Here $f_{\pi} \sim \sqrt{N_{c}}$ is the pion decay constant, $g_{A}\sim N_{c}^{1}$ is the axial coupling constant, and $\Delta \equiv (M_{\Delta} - M_{N})/\Lambda_{QCD} \sim 1/N_{c}$ is the nucleon-delta mass splitting.  The ratio in Eq.~\eqref{eq:MagicRatio} is constructed to make the low-energy constants cancel, so that it should take the same numerical value in any theory with baryons and the same pattern of chiral symmetry breaking as in QCD.  Of course, this assumes that a particular model in question does not suffer from the issues discussed in the introduction to do with the an accidental (from the perspective of \XPT ) suppression of some of these low-energy constants.  If such suppressions are present and come in an unfortunate pattern, the long-range physics might end up being qualitatively different.

The quantization of the soliton system is standard\cite{Adkins:1983ya}, and is discussed at length in this notation in Ref.~\cite{Panico:2008it}.  For our purposes, the important outputs of this analysis are the following identities for the matrix elements of the collective coordinates on the subspace of nucleon states:
\begin{align}
\langle \Tr U \sigma^{b} U^{\dag} \sigma^{a} \rangle &= - \frac{8}{3} S^{b} I^{a} , \\
\langle \Tr U \sigma^{b} \hat{x}_{b}(\vec{k}\cdot \hat{x} U^{\dag} \sigma^{a} \rangle &= - \frac{2}{3 \Lambda} I^{a} , \\
\end{align}
where $S^{a}$ and $I^{a}$ are the expectation values of spin and isospin respectively.

Armed with these identities, we can plug the expressions for the isoscalar and isovector currents in Eq.~\eqref{eq:CurrentDefinitions} into the definitions of the position-space electromagnetic form factors in Eq.~\eqref{eq:FormFactorDefinition}, and thence obtain simple expressions for the electromagnetic form factors in terms of the fields parametrizing the symmetry-reduced ansatz in Eqs.~\eqref{eq:StaticAnsatz},~\eqref{eq:TimeDependentAnsatz}.  The result of these manipulations is
\begin{align}
\label{eq:HolographicFormFactors}
\tilde{G}_E^{I=0}(r) &= -{4\over N_c} \kappa \left[ k(z)\partial_z s  \right]^{z_{\mathrm{uv}}}_{-z_{\mathrm{uv}}}, \\
\tilde{G}_M^{I=0}(r)&= -{2\over 3N_c\Lambda} \kappa \left[ r k(z)\partial_z Q~ \right]^{z_{\mathrm{uv}}}_{-z_{\mathrm{uv}}}, \\
\tilde{G}_E^{I=1}(r) &= {2\over 3\Lambda}\kappa \left[k(z)(\partial_z v-2 (\partial_z\chi_2- A_z \chi_1)) \right]^{z_{\mathrm{uv}}}_{-z_{\mathrm{uv}}}, \\
\tilde{G}_M^{I=1}(r) &= -{4\over9}\kappa \left[ k(z)(\partial_z \phi_2 - A_z \phi_1)\right]^{z_{\mathrm{uv}}}_{-z_{\mathrm{uv}}},
\end{align}

Now we are basically done, since we have everything we need to check the model-independent relation Eq.~\eqref{eq:MagicRatio}. Plugging in the large $r$ solutions in Eqs.~\eqref{eq:StaticSolutions},~\eqref{eq:TimeDependentSolutions}  into Eq.~\eqref{eq:HolographicFormFactors}, and extracting the leading terms at large $r$ and $z_{\mathrm{uv}}$, we get
\begin{align}
\label{eq:FormFactorResults}
\tilde{G}_E^{I=0} &\to \frac{32 \kappa z_{\mathrm{uv}}^3 \beta ^3 \gamma }{N_c r^9} , \\
\tilde{G}_M^{I=0}(r) & \to -\frac{16 \kappa z_{\mathrm{uv}}^3 \beta ^3 \gamma }{3 N_c r^7 \Lambda } , \nnb\\
\tilde{G}_E^{I=1}(r) & \to -\frac{16 \kappa z_{\mathrm{uv}}^2 \beta ^2}{3 \pi  r^4 \Lambda } , \nnb \\
\tilde{G}_M^{I=1}(r) & \to \frac{16 \kappa z_{\mathrm{uv}}^2 \beta ^2}{9 \pi  r^4} \nnb .
\end{align}
Note that in all of these expressions $\beta$ enters in the combination $\beta z_{\mathrm{uv}}$.  The form factors are observables, and so cannot depend on a `UV' cutoff or any choice of renormalization scheme in the gravity dual.  However, the detailed expressions of physical observables in terms of model parameters can certainly depend on the choice of scheme.  There are two generic ways these expressions may become independent from the cutoff.  The first is if $\beta$, which is determined by the global solution to the boundary value problem for the $Q=1$ soliton, scales as $\beta \sim 1/z_{\mathrm{uv}}$ when $z_{\mathrm{uv}} \gg 1$.  Alternatively, $\beta$ may scale differently, and there may be contributions from $S_{b}$ to these relations that we did not explicitly write. To work this out completely, one could for instance verify that when Eqs.~\eqref{eq:FormFactorResults} are rewritten in terms of  $f_{\pi},g_{A}$, and $\Delta$, all scheme and cutoff dependence disappears, so long as  $f_{\pi},g_{A}$, and $\Delta$ are themselves calculated using the same cutoff we used above.

Fortunately, for our main goal of calculating Eq.~\eqref{eq:MagicRatio}, these difficult issues do not have to be resolved. Assembling the ratio, we find that
\begin{align}\label{eq:SSmodelRatio}
\lim_{r\to\infty} r^{2} \frac{\tilde{G}_{E}^{I=0}\tilde{G}_{E}^{I=1}}{\tilde{G}_{M}^{I=0}\tilde{G}_{M}^{I=1}} = 18,
\end{align}
in the Sakai-Sugimoto model, as advertised.  As we have been foreshadowing, all of the model parameters and the associated subtleties cancel from the ratio Eq.~\eqref{eq:SSmodelRatio}, as they must in any model that consistently implements the  structure of baryon-pion physics expected from large $N_{c}$ chiral perturbation theory.  The reassuring implication is that the structure of the long-range physics of baryons in the Sakai-Sugimoto model is the same as in large $N_{c}$ QCD.

%%%%%%%%%%%%%%%%%%%%%
\section{Discussion}
\label{sec:discussion}
%%%%%%%%%%%%%%%%%%%%%
Our results provide strong evidence that baryons in the Sakai-Sugimoto model obey the model-independent relation Eq.~\eqref{eq:MagicRatio}. To see this, we focused on the behavior of the soliton solution of the model at large distances $r$.  Armed with asymptotic solutions of the equations of motion obtained in power series expansion in $1/r$ at large $r$, we evaluated the large $r$ contributions to the electromagnetic form factors of the proton. The form factors behaved in precisely the way expected from \XPT, and as a result the large $r$ form factors obeyed Eq.~\eqref{eq:MagicRatio}. So the issues of the Sakai-Sugimoto model appearing to fail to meet the expectations of \XPT\ raised in Ref.~\cite{Cherman:2009gb} seem to be connected to subtle issues with the applicability of the standard approach in the literature of using the flat-space instantons approximation, rather than a deep problem with the model itself.  

It remains a very interesting task to understand exactly why the standard flat-space instanton approach gives misleading results for the observables discussed here, and to clarify the circumstances in which the standard approach should be reliable, since it leads to apparently reasonable  predictions in many cases.  Here we note a couple of tentative possible reasons for the discrepancies between our results and those in the literature.  The first one has to do with the standard approach in computations in the Sakai-Sugimoto model, where $\zuv$ is taken to infinity from the start of the calculation.  As we saw in the direct construction of the large $r$ solutions, sending $\zuv \to \infty$ too early can cause one to misidentify the asymptotic structure of the soliton fields.  Another possible issue concerns the usual approach for extracting the behavior of the currents from the flat-space instanton solution.  For this it is necessary to know the behavior of the fields at large $z$, but there the flat-space instanton approximation cannot be trusted.  The standard approach is to argue that the soliton solution linearizes already at $z \ll 1$, and the flat-space instanton solution is matched to the solutions of the linearized equation of motion for $z \gg1$.  But as we saw in the direct solution at large $r$, while the system certainly linearizes in the sense that an order-by-order solution is possible, the non-linear terms in the full equations of motion play an important role in determining the solutions.  This suggests that the task of connecting the flat-space instanton to the behavior of the soliton fields at large $z$ may be rather subtle.

There are many important directions for future work.  Perhaps the most urgent task is to find a numerical $Q=1$ solution to the equations of motion, and match it both to our large $r$ solution and to the $r, z \ll \lambda^{-1/2}$ flat-space instanton solutions in the literature.   Among other things, this would greatly help in understanding the connections between the approach taken here and the flat-space instanton approximation.  

Another urgent and possibly related task is to undertake a systematic study of holographic renormalization for the Sakai-Sugimoto model.  Thanks to the special nature of the observable we were aiming at [Eq.~\eqref{eq:MagicRatio}], we were able to get away with being somewhat cavalier in our treatment of such issues in our current work, but the problem clearly cries out for a careful treatment.  A detailed understanding of holographic renormalization and of the boundary counterterm action $S_{b}$ may be important both for constructing numerical soliton solutions, and for understanding the domain of validity and proper interpretation of the previous calculations in the literature, both in the meson and baryon sectors.  

Finally, there may be some phenomenological implications of our results.  It would be interesting to revisit the many baryonic observables already computed in the Sakai-Sugimoto model, and see whether going beyond the flat-space instanton approximation improves the phenomenological match to real-world data.  It is possible that such an approach may shed some light on the large-distance interactions between baryons in the Sakai-Sugimoto model, which are known to be difficult to accurately capture with the flat-space instanton approximation\cite{Hashimoto:2009ys,Hashimoto:2010je,Hashimoto:2009as,Hashimoto:2010ue,Kim:2009sr,Kaplunovsky:2010eh,Dymarsky:2010ci}.  As another direction, it may  be interesting to consider the implications of our results for the description of baryons with valence strange quarks away from the $SU(3)$ flavor limit, which are most generally approached using the bound-state approach \cite{Callan:1985hy,Callan:1987xt} in the context of the Skyrme model. A trial study in this direction in the Sakai-Sugimoto model has been performed in Ref.~\cite{Ishii:2010ib}, where flat-space instanton approximation was adopted. Since it seems that contributions of pion clouds may be significant for the bound-state approach, it will be interesting to reexamine these calculations in light of the present results.

%%%%%%%%%%%%%%%%%
\section*{Acknowledgements}
%%%%%%%%%%%%%%%%%

We are very grateful to Nick Manton for collaboration at the early stages of this work, and for many enlightening discussions and encouragement throughout.  We also thank Tom Cohen, Koji Hashimoto, Sungjay Lee, Tadakatsu Sakai, and Shigeki Sugimoto for very helpful conversations.  A.C. thanks the Galileo Galilei Institute for hospitality and the INFN for partial support during the program on ``Large N Gauge Theories'', where part of this work was completed.

\bibliography{bib18} 

\providecommand{\href}[2]{#2}\begingroup\raggedright\begin{thebibliography}{10}

\bibitem{Maldacena:1997re}
J.~M. Maldacena, {\it {The Large N limit of superconformal field theories and
  supergravity}},  {\em Adv.Theor.Math.Phys.} {\bf 2} (1998) 231--252,
  [\href{http://xxx.lanl.gov/abs/hep-th/9711200}{{\tt hep-th/9711200}}].

\bibitem{Witten:1998xy}
E.~Witten, {\it {Baryons and branes in anti de Sitter space}},  {\em JHEP} {\bf
  07} (1998) 006, [\href{http://xxx.lanl.gov/abs/hep-th/9805112}{{\tt
  hep-th/9805112}}].

\bibitem{Gubser:1998bc}
S.~S. Gubser, I.~R. Klebanov, and A.~M. Polyakov, {\it {Gauge theory
  correlators from non-critical string theory}},  {\em Phys. Lett.} {\bf B428}
  (1998) 105--114, [\href{http://xxx.lanl.gov/abs/hep-th/9802109}{{\tt
  hep-th/9802109}}].

\bibitem{Sakai:2004cn}
T.~Sakai and S.~Sugimoto, {\it {Low energy hadron physics in holographic QCD}},
   {\em Prog. Theor. Phys.} {\bf 113} (2005) 843--882,
  [\href{http://xxx.lanl.gov/abs/hep-th/0412141}{{\tt hep-th/0412141}}].

\bibitem{Sakai:2005yt}
T.~Sakai and S.~Sugimoto, {\it {More on a holographic dual of QCD}},  {\em
  Prog. Theor. Phys.} {\bf 114} (2005) 1083--1118,
  [\href{http://xxx.lanl.gov/abs/hep-th/0507073}{{\tt hep-th/0507073}}].

\bibitem{tHooft:1973jz}
G.~t~Hooft, {\it A planar diagram theory for strong interactions},  {\em Nucl.
  Phys.} {\bf B72} (1974).

\bibitem{Witten:1979kh}
E.~Witten, {\it {Baryons in the 1/n Expansion}},  {\em Nucl. Phys.} {\bf B160}
  (1979) 57.

\bibitem{Hong:2007kx}
D.~K. Hong, M.~Rho, H.-U. Yee, and P.~Yi, {\it {Chiral dynamics of baryons from
  string theory}},  {\em Phys. Rev.} {\bf D76} (2007) 061901,
  [\href{http://xxx.lanl.gov/abs/hep-th/0701276}{{\tt hep-th/0701276}}].

\bibitem{Hata:2007mb}
H.~Hata, T.~Sakai, S.~Sugimoto, and S.~Yamato, {\it {Baryons from instantons in
  holographic QCD}},  {\em Prog. Theor. Phys.} {\bf 117} (2007) 1157,
  [\href{http://xxx.lanl.gov/abs/hep-th/0701280}{{\tt hep-th/0701280}}].

\bibitem{Hong:2007ay}
D.~K. Hong, M.~Rho, H.-U. Yee, and P.~Yi, {\it {Dynamics of Baryons from String
  Theory and Vector Dominance}},  {\em JHEP} {\bf 09} (2007) 063,
  [\href{http://xxx.lanl.gov/abs/0705.2632}{{\tt arXiv:0705.2632}}].

\bibitem{Hong:2007dq}
D.~K. Hong, M.~Rho, H.-U. Yee, and P.~Yi, {\it {Nucleon form-factors and hidden
  symmetry in holographic QCD}},  {\em Phys.Rev.} {\bf D77} (2008) 014030,
  [\href{http://xxx.lanl.gov/abs/0710.4615}{{\tt arXiv:0710.4615}}].

\bibitem{Park:2008sp}
J.~Park and P.~Yi, {\it {A Holographic QCD and Excited Baryons from String
  Theory}},  {\em JHEP} {\bf 0806} (2008) 011,
  [\href{http://xxx.lanl.gov/abs/0804.2926}{{\tt arXiv:0804.2926}}].

\bibitem{Hashimoto:2008zw}
K.~Hashimoto, T.~Sakai, and S.~Sugimoto, {\it {Holographic Baryons : Static
  Properties and Form Factors from Gauge/String Duality}},  {\em Prog. Theor.
  Phys.} {\bf 120} (2008) 1093--1137,
  [\href{http://xxx.lanl.gov/abs/0806.3122}{{\tt arXiv:0806.3122}}].

\bibitem{Kim:2008pw}
K.-Y. Kim and I.~Zahed, {\it {Electromagnetic Baryon Form Factors from
  Holographic QCD}},  {\em JHEP} {\bf 09} (2008) 007,
  [\href{http://xxx.lanl.gov/abs/0807.0033}{{\tt arXiv:0807.0033}}].

\bibitem{Cherman:2009gb}
A.~Cherman, T.~D. Cohen, and M.~Nielsen, {\it {Model Independent Tests of
  Skyrmions and Their Holographic Cousins}},  {\em Phys. Rev. Lett.} {\bf 103}
  (2009) 022001, [\href{http://xxx.lanl.gov/abs/0903.2662}{{\tt
  arXiv:0903.2662}}].

\bibitem{Domenech:2010aq}
O.~Domenech, G.~Panico, and A.~Wulzer, {\it {Massive Pions, Anomalies and
  Baryons in Holographic QCD}},  {\em Nucl.Phys.} {\bf A853} (2011) 97--123,
  [\href{http://xxx.lanl.gov/abs/1009.0711}{{\tt arXiv:1009.0711}}].

\bibitem{Cohen:1989qk}
T.~D. Cohen, {\it {LARGE N(c) QCD, COMPOSITE NUCLEONS AND THE DIRAC SEA}},
  {\em Phys.Rev.Lett.} {\bf 62} (1989) 3027.

\bibitem{Kiritsis:1988qz}
E.~Kiritsis and R.~Seki, {\it {YUKAWA THEORIES AS EFFECTIVE THEORIES OF QUANTUM
  CHROMODYNAMICS FOR A LARGE NUMBER OF COLORS}},  {\em Phys.Rev.Lett.} {\bf 63}
  (1989) 953.

\bibitem{Arnold:1990fp}
P.~B. Arnold and M.~P. Mattis, {\it {SUMMING GRAPHS IN LARGE N(c) QUANTUM
  HADRODYNAMICS}},  {\em Phys.Rev.Lett.} {\bf 65} (1990) 831--834.

\bibitem{Pomarol:2007kr}
A.~Pomarol and A.~Wulzer, {\it {Stable skyrmions from extra dimensions}},  {\em
  JHEP} {\bf 03} (2008) 051--051,
  [\href{http://xxx.lanl.gov/abs/0712.3276}{{\tt arXiv:0712.3276}}].

\bibitem{Pomarol:2008aa}
A.~Pomarol and A.~Wulzer, {\it {Baryon Physics in Holographic QCD}},  {\em
  Nucl. Phys.} {\bf B809} (2009) 347--361,
  [\href{http://xxx.lanl.gov/abs/0807.0316}{{\tt arXiv:0807.0316}}].

\bibitem{Panico:2008it}
G.~Panico and A.~Wulzer, {\it {Nucleon Form Factors from 5D Skyrmions}},  {\em
  Nucl. Phys.} {\bf A825} (2009) 91--114,
  [\href{http://xxx.lanl.gov/abs/0811.2211}{{\tt arXiv:0811.2211}}].

\bibitem{Witten:1998zw}
E.~Witten, {\it {Anti-de Sitter space, thermal phase transition, and
  confinement in gauge theories}},  {\em Adv. Theor. Math. Phys.} {\bf 2}
  (1998) 505--532, [\href{http://xxx.lanl.gov/abs/hep-th/9803131}{{\tt
  hep-th/9803131}}].

\bibitem{Mandal:2011ws}
G.~Mandal and T.~Morita, {\it {Gregory-Laflamme as the
  confinement/deconfinement transition in holographic QCD}},
  \href{http://xxx.lanl.gov/abs/1107.4048}{{\tt arXiv:1107.4048}}.

\bibitem{Karch:2002sh}
A.~Karch and E.~Katz, {\it {Adding flavor to AdS/CFT}},  {\em JHEP} {\bf 06}
  (2002) 043, [\href{http://xxx.lanl.gov/abs/hep-th/0205236}{{\tt
  hep-th/0205236}}].

\bibitem{Itzhaki:1998dd}
N.~Itzhaki, J.~M. Maldacena, J.~Sonnenschein, and S.~Yankielowicz, {\it
  {Supergravity and the large N limit of theories with sixteen supercharges}},
  {\em Phys. Rev.} {\bf D58} (1998) 046004,
  [\href{http://xxx.lanl.gov/abs/hep-th/9802042}{{\tt hep-th/9802042}}].

\bibitem{Hashimoto:2010je}
K.~Hashimoto, N.~Iizuka, and P.~Yi, {\it {A Matrix Model for Baryons and
  Nuclear Forces}},  {\em JHEP} {\bf 10} (2010) 003,
  [\href{http://xxx.lanl.gov/abs/1003.4988}{{\tt arXiv:1003.4988}}].

\bibitem{Skenderis:2002wp}
K.~Skenderis, {\it {Lecture notes on holographic renormalization}},  {\em
  Class. Quant. Grav.} {\bf 19} (2002) 5849--5876,
  [\href{http://xxx.lanl.gov/abs/hep-th/0209067}{{\tt hep-th/0209067}}].

\bibitem{Kanitscheider:2008kd}
I.~Kanitscheider, K.~Skenderis, and M.~Taylor, {\it {Precision holography for
  non-conformal branes}},  {\em JHEP} {\bf 09} (2008) 094,
  [\href{http://xxx.lanl.gov/abs/0807.3324}{{\tt arXiv:0807.3324}}].

\bibitem{Benincasa:2009ze}
P.~Benincasa, {\it {A Note on Holographic Renormalization of Probe D-Branes}},
  \href{http://xxx.lanl.gov/abs/0903.4356}{{\tt arXiv:0903.4356}}.

\bibitem{Papadimitriou:2010as}
I.~Papadimitriou, {\it {Holographic renormalization as a canonical
  transformation}},  {\em JHEP} {\bf 11} (2010) 014,
  [\href{http://xxx.lanl.gov/abs/1007.4592}{{\tt arXiv:1007.4592}}].

\bibitem{Papadimitriou:2011qb}
I.~Papadimitriou, {\it {Holographic Renormalization of general dilaton-axion
  gravity}},  {\em JHEP} {\bf 08} (2011) 119,
  [\href{http://xxx.lanl.gov/abs/1106.4826}{{\tt arXiv:1106.4826}}].

\bibitem{Hashimoto:2005qh}
K.~Hashimoto and S.~Terashima, {\it {ADHM is tachyon condensation}},  {\em
  JHEP} {\bf 02} (2006) 018,
  [\href{http://xxx.lanl.gov/abs/hep-th/0511297}{{\tt hep-th/0511297}}].

\bibitem{Collie:2008vc}
B.~Collie and D.~Tong, {\it {Instantons, Fermions and Chern-Simons Terms}},
  {\em JHEP} {\bf 07} (2008) 015,
  [\href{http://xxx.lanl.gov/abs/0804.1772}{{\tt arXiv:0804.1772}}].

\bibitem{Kim:2008kn}
S.~Kim, K.-M. Lee, and S.~Lee, {\it {Dyonic Instantons in 5-dim Yang-Mills
  Chern-Simons Theories}},  {\em JHEP} {\bf 0808} (2008) 064,
  [\href{http://xxx.lanl.gov/abs/0804.1207}{{\tt arXiv:0804.1207}}].

\bibitem{Forgacs:1979zs}
P.~Forgacs and N.~S. Manton, {\it {Space-Time Symmetries in Gauge Theories}},
  {\em Commun. Math. Phys.} {\bf 72} (1980) 15.

\bibitem{Witten:1976ck}
E.~Witten, {\it {Some exact multipseudoparticle solutions of classical
  Yang-Mills theory}},  {\em Phys. Rev. Lett.} {\bf 38} (1977) 121.

\bibitem{Adkins:1983ya}
G.~S. Adkins, C.~R. Nappi, and E.~Witten, {\it {Static Properties of Nucleons
  in the Skyrme Model}},  {\em Nucl. Phys.} {\bf B228} (1983) 552.

\bibitem{Hashimoto:2009ys}
K.~Hashimoto, T.~Sakai, and S.~Sugimoto, {\it {Nuclear Force from String
  Theory}},  {\em Prog. Theor. Phys.} {\bf 122} (2009) 427--476,
  [\href{http://xxx.lanl.gov/abs/0901.4449}{{\tt arXiv:0901.4449}}].

\bibitem{Hashimoto:2009as}
K.~Hashimoto, N.~Iizuka, and T.~Nakatsukasa, {\it {N-Body Nuclear Forces at
  Short Distances in Holographic QCD}},  {\em Phys. Rev.} {\bf D81} (2010)
  106003, [\href{http://xxx.lanl.gov/abs/0911.1035}{{\tt arXiv:0911.1035}}].

\bibitem{Hashimoto:2010ue}
K.~Hashimoto and N.~Iizuka, {\it {Three-Body Nuclear Forces from a Matrix
  Model}},  {\em JHEP} {\bf 11} (2010) 058,
  [\href{http://xxx.lanl.gov/abs/1005.4412}{{\tt arXiv:1005.4412}}].

\bibitem{Kim:2009sr}
Y.~Kim, S.~Lee, and P.~Yi, {\it {Holographic Deuteron and Nucleon-Nucleon
  Potential}},  {\em JHEP} {\bf 04} (2009) 086,
  [\href{http://xxx.lanl.gov/abs/0902.4048}{{\tt arXiv:0902.4048}}].

\bibitem{Kaplunovsky:2010eh}
V.~Kaplunovsky and J.~Sonnenschein, {\it {Searching for an Attractive Force in
  Holographic Nuclear Physics}},  {\em JHEP} {\bf 05} (2011) 058,
  [\href{http://xxx.lanl.gov/abs/1003.2621}{{\tt arXiv:1003.2621}}].

\bibitem{Dymarsky:2010ci}
A.~Dymarsky, D.~Melnikov, and J.~Sonnenschein, {\it {Attractive Holographic
  Baryons}},  {\em JHEP} {\bf 06} (2011) 145,
  [\href{http://xxx.lanl.gov/abs/1012.1616}{{\tt arXiv:1012.1616}}].

\bibitem{Callan:1985hy}
J.~Callan, Curtis~G. and I.~R. Klebanov, {\it {Bound State Approach to
  Strangeness in the Skyrme Model}},  {\em Nucl. Phys.} {\bf B262} (1985) 365.

\bibitem{Callan:1987xt}
J.~Callan, Curtis~G., K.~Hornbostel, and I.~R. Klebanov, {\it {Baryon Masses in
  the Bound State Approach to Strangeness in the Skyrme Model}},  {\em
  Phys.Lett.} {\bf B202} (1988) 269.

\bibitem{Ishii:2010ib}
T.~Ishii, {\it {Toward Bound-State Approach to Strangeness in Holographic
  QCD}},  {\em Phys.Lett.} {\bf B695} (2011) 392--396,
  [\href{http://xxx.lanl.gov/abs/1009.0986}{{\tt arXiv:1009.0986}}].

\end{thebibliography}\endgroup

\end{document}